\documentclass[namedreferences]{SolarPhysics}
\usepackage[optionalrh]{spr-sola-addons}
\usepackage{graphicx}
\begin{document}
\begin{article}
\begin{opening}
\title{Helmet Streamers with Triple Structures: \\
Simulations of resistive dynamics.}
\runningtitle{Helmet Streamers}
\author{Thomas \surname{Wiegelmann}}
\institute{Max Planck Institut f\"ur Aeronomie, Max Planck
Stra\ss e 2,
D-37191 Katlenburg-Lindau, Germany}
\author{Karl \surname{Schindler}}
\institute{Institut f\"ur Theoretische Physik IV, Ruhr-Universit\"at Bochum,
D-44780 Bochum, Germany}
\author{Thomas \surname{Neukirch}}
\institute{School of Mathematical and Computational Sciences,
University of St. Andrews,
St. Andrews, Scotland}
\runningauthor{Wiegelmann et al.}
\date{Bibliographic Code: 2000SoPh..191..391W}

\begin{abstract}
Recent observations of the solar corona with the
LASCO coronagraph on board of the SOHO spacecraft have revealed the
occurrence of triple helmet streamers even during solar minimum, which
occasionally go unstable and give rise to large coronal mass ejections.
There are also indications that the slow solar wind is
either a combination of a quasi-stationary
flow and a highly fluctuating component or may  even be caused completely by
many small eruptions or instabilities.
As a first step we recently presented
an analytical method to
calculate simple two-dimensional stationary models
of triple helmet streamer configurations. In the present contribution
we use the equations of time-dependent resistive
magnetohydrodynamics to investigate the stability and the
dynamical behaviour of these
configurations. We  particularly focus on the possible differences
between the dynamics of single isolated streamers
and triple streamers and  on the way
in which magnetic reconnection initiates both small scale
and large scale dynamical behaviour of the streamers.
Our results indicate that small eruptions at the helmet streamer cusp may
incessantly
accelerate  small amounts of plasma without significant changes of
the equilibrium configuration and might
thus contribute to the non-stationary slow solar wind. On
larger time and length scales,
large coronal eruptions can occur as a consequence
of large scale magnetic reconnection events
inside the streamer configuration.
Our results also show that
triple streamers are usually more stable than a single streamer.
\end{abstract}

\end{opening}

\section{Introduction}
Recent observations of the corona
with the
LASCO coronagraph
\cite{schwenn97}
on board of the SOHO spacecraft showed that the corona can be highly
structured even during the solar activity minimum. The observations
revealed a triple structure of the streamer belt which was existent
for several consecutive days.
The observations further showed
that these triple structures occasionally go unstable leading to a
seemingly new and extraordinarily huge kind of coronal mass ejection
(global CMEs). Natural questions arising from these observations are
whether the helmet streamer triple structure is directly
connected with or responsible
for the occurrence of global CMEs and what is the physical mechanism of their formation.

The structure of helmet streamers and their stability has been
studied both observationally and theoretically for a long time
(e.g. \opencite{pneuman:kopp71};
      \opencite{cuperman:etal90};
      \opencite{cuperman:etal92};
      \opencite{koutchmy:livshits92};
      \opencite{wang:etal93};
      \opencite{cuperman:etal95};
      \opencite{wu:etal95};
      \opencite{bavassano:etal97};
      \opencite{noci};
\opencite{hundhausen99}).
There seems to be a natural association of
helmet streamers with coronal eruptions and coronal streamers are
assumed to be the source region of the slow solar wind. The
traditional view towards the origin of the slow solar wind is that it
is a more or less
stationary plasma flow on open field lines around the
closed field lines of a helmet streamer. Recent observations
(e.g. \opencite{habbal98};
 \opencite{noci})
challenge this
traditional view and indicate that the slow solar wind is
non-stationary and seems to be
produced and accelerated by small eruptions in
the helmet streamer stalk above the cusp. This acceleration
process of the slow solar wind has been compared with the rise of
smoke above a burning candle (Schwenn, private communication).

Pre-SOHO observations of multiple streamer configurations during
the maximum phase of solar activity and of the multiple current sheet
structure of the heliospheric plasma sheet
\cite{crooker:etal93,woo:etal95} have initiated
several studies of the dynamics and stability of
multiple current sheets with variations in only one spatial dimension
\cite{otto:birk92,yan:etal94,
dahlburg:karpen95,birk:etal97,wang:etal97}.
\inlinecite{einaudi:etal99} have recently presented a model for the
generation of the non-stationary slow solar wind based on linear
and non-linear stability calculations for a single one-dimensional
current sheet with field-aligned flow.

All these models do, however,
in a strict sense only apply to the streamer stalk, i.e.\ to the
open field lines of the heliospheric current sheet. Here we aim to
investigate both closed and parts of the adjacent
open field line regions.

Models of multiple arcade and loop structures have been investigated before
by e.g.\ \inlinecite{mikic:etal88}, \inlinecite{biskamp:welter89}
and most recently by \inlinecite{antiochos:etal99}.
Our model differs from these models basically by the possibility
of having a flexible
analytical initial condition for the time-dependent
calculations allowing the investigation of different types
of structures.

As a first step towards improving the theoretical understanding of
the above mentioned phenomena in triple streamer configurations,
we have calculated analytic two-dimensional static models
of triple helmet streamers
in a previous paper (\opencite{paper1};
further cited
as {\it paper I}).
The aim of the present paper is to undertake the next step
in this investigation and to
study the stability of the stationary state helmet streamer
configurations calculated in {\it paper I}. We will
carry this out with the help of
time-dependent numerical experiments
using the equations of resistive magnetohydrodynamics.

The outline of the paper is as follows. In Section \ref{basics}
we discuss the basic equations and briefly describe
our numerical method. Section
\ref{model} outlines our main model
assumptions. In Section \ref{results} we present the results of
numerical experiments and
in Section \ref{conclusions} we discuss our results and give an
outlook on future work.

\section{Basic Equations and Numerical Method}
\label{basics}

We use the equations of time-dependent resistive
magnetohydrodynamics (MHD) to describe the coronal plasma.
For a discussion concerning the neglect of gravity see sect.
\ref{init}.
\begin{eqnarray}
  -\rho {\bf v}\cdot\nabla  {\bf v}  -\nabla P + {\bf j} \times {\bf B} &=&  \rho \frac{\partial \bf v}{\partial t}    \label{gl1} \\
     - \nabla\cdot (\rho{\bf v})&=& \frac{\partial \rho}{\partial t}                           \label{gl2} \\
{\bf E} + {\bf v} \times {\bf B} &=&  \eta {\bf j}        \label{gl3}  \\
\nabla \cdot {\bf B} &=&0                                 \label{gl4} \\
{\bf j} &=& \frac{1}{\mu_0} \nabla \times {\bf B}
\label{gl5} \\
P &=& \rho R T  \label{gl6} \\
\frac{\partial {\bf B}}{\partial t} &=&-\nabla \times {\bf E}
\label{gl7} \\
\frac{\partial P}{\partial t} +  \nabla \cdot (P {\bf v})
+(\gamma-1)P \nabla \cdot {\bf v} &=& (\gamma-1) \eta j^2 \label{gl8}
\end{eqnarray}
Here, $P$ stands for the plasma pressure,
$\bf B$ for the magnetic field, $\rho$ for the
plasma density,
$\bf v$ for the plasma velocity,
$\bf E$ for the electric field,
$R$ is the gas constant,
$T$ the temperature, ${\bf j}$ the current density, $\eta$ the
resistivity, $\gamma$ the adiabatic index
 and $\mu_0$ the vacuum permeability. We normalize
the magnetic field
by a typical value $B_0$, the plasma pressure $P$  by
$B_0^2/\mu_0$,
the mass density $\rho$ by $\rho_0=B_0^2/\mu_0 \mbox{R} T$,
the length $L$ by a solar radius and the current density
by $B_0/\mu_0\,L$, the plasma bulk velocity by the Alfven velocity $v_A$, the
electric field by $\frac{B_0^2}{\sqrt{\rho_0}}$, the time by the Alfven time and the
resistivity
$\eta$ by $\mu_0 L v_A$.

The time-dependent MHD equations are highly non-linear and
generally are solved numerically. Our code uses an explicit
finite difference scheme and
is described in detail by \inlinecite{dissjd} and \inlinecite{disslr}.
It has been successfully applied
to several astrophysical problems (e.g. \opencite{dissjd,lrtn,disslr}).

One of the fundamental problems of any numerical experiment
based on the non-ideal MHD equations is the strength of the
resistive term. It is well-known that the level of disspation in the solar
corona is very small (magnetic Reynolds number $R_m \simeq 10^{12}$ for
Spitzer resistivity or at most a few orders magnitude lower if one
allows for anomalous resistivity of some kind). This means that either
the spatial resolution of the numerical code has to be large enough to
resolve the small spatial length scales associated with the
small disspative terms or that one has to use unrealistically large values
for the dissipative coefficients. This second approach is
based on the assumption that the basic dynamics of resistive processes
is not changed as long as $R_m \gg 1$.
Since the first approach cannot be carried out
due to the limited capacity of modern computers,
we use the second approach.

Because of the nonzero resistivity $\eta$ the MHD equations do not
allow static solutions in a strict sense because of
magnetic diffusion. However, it is well known that in most astrophysical
plasmas the resistivity and thus the diffusive  terms are very small.
The diffusive time scale is in fact typically much (some orders of magnitude)
larger than the dynamic time scale. As the dynamic time scale for static
equilibria we take the time scale on which instabilities occur.
On this time scale the magnetic diffusivity can be neglected for
magnetohydrostatic equilibria in the absence of thin current sheets. If
thin current sheets are present in the equilibria, magnetic diffusion
on time scales short compared with typical macroscopic scales of the
system might become important and lead to magnetic reconnection.

\section{Model Assumptions}
\label{model}

\subsection{The initial conditions}
\label{init}
We use the analytical stationary equilibria calculated in
{\it paper I}
as initial conditions for the time-dependent
numerical MHD experiments.
The LASCO observations (\citeauthor{schwenn97},
\citeyear{schwenn97}) show that the streamer belt
is fairly extended in azimuth. Thus we restrict
our calculations to  two dimensions
as a first step.

As pointed out in {\it paper I}
we do not include plasma flow in the stationary states of our
helmet streamer configurations. The observations
(e.g. \opencite{habbal98}) give
strong evidence that a stationary slow solar wind does not
exist. These observations further suggest that continuous small
instabilities and eruptions lead
to the formation
of a non-stationary slow solar wind. It is one aim of this paper
to investigate possible mechanisms which could
lead to such a behaviour
within the framework of our model.

We simplify the calculations by using
Cartesian geometry and neglecting
solar gravity.
Of course, both gravity and the geometry used
will have a quantitative influence on the dynamics of the
system. However, the main interest of this work is to identify
the possible mechanism of acceleration of the slow solar wind
and the instabilities leading to coronal
mass ejections in triple streamer configurations.
We expect that the  question whether an instability
occurs or not will be dominated more by the topology of magnetic fields and
less by the inclusion of gravity or by the geometry used
for the calculation. The price we
have to pay for making these assumptions is of course
that we cannot expect a
quantitative agreement of, e.g. plasma flow velocity or plasmoid
velocity with the observed data. In any case the present idealized
study seems necessary as a first step towards a more realistic description.

Before we investigate the resistive dynamics of
helmet streamers we first have to investigate the
equilibria concerning their stability within the framework of ideal
MHD. Only if no significant changes of the configurations occur
over many Alfv\'en crossing times under ideal conditions,
investigations of instabilities
in the framework of resistive MHD are meaningful.
We investigated all configurations presented in {\it paper I},
table I and table II, numerically in the framework of ideal MHD and found that
the changes in the values of kinetic energy, total energy, magnetic
energy, thermal energy and total mass is less than one percent in all
cases for $140$ Alfv\'en crossing times
and thus the configurations may be regarded as ideally stable. Thus
instabilities with significant plasma flow and eruptions are only
to be expected if resistive effects are included. We remark that
this result can be expected since the slender streamer configurations
are rather close to one dimensional structures, which are known to be stable in
ideal MHD  \cite{ks83}.

\subsection{Resistivity Profile}
\label{resistivity}

Of crucial importance for
numerical experiments of resistive instabilities  are
the assumptions made for the dissipative terms. As we have already
mentioned,
the magnetic Reynolds number due to collisions
in the coronal plasma is extremely
large and thus the electric resistivity can usually
be considered as being approximately zero.
However, in localized regions with a high electric
current density, plasma micro-instabilities may occur leading to
an anomalous resistivity which can be several orders of magnetitude
larger than the collisional resistivity. As the exact mechanisms for
the generation of anomalous resistivity are still not fully known, we
 use ad hoc resistivity models.
Fortunately, it is known
from investigations of the Earth's magnetotail
\cite{dissao,otto:etal90} that different resistivity
profiles do not influence the qualitative results significantly,
while details may be different. We have
investigated three different resistivity models:
\begin{enumerate}
\item spatially constant and time-independent resistivity,
\item spatially localized and time-independent resistivity,
\item current dependent resistivity.
\end{enumerate}
\begin{figure}
\includegraphics[height=13cm,width=10cm]{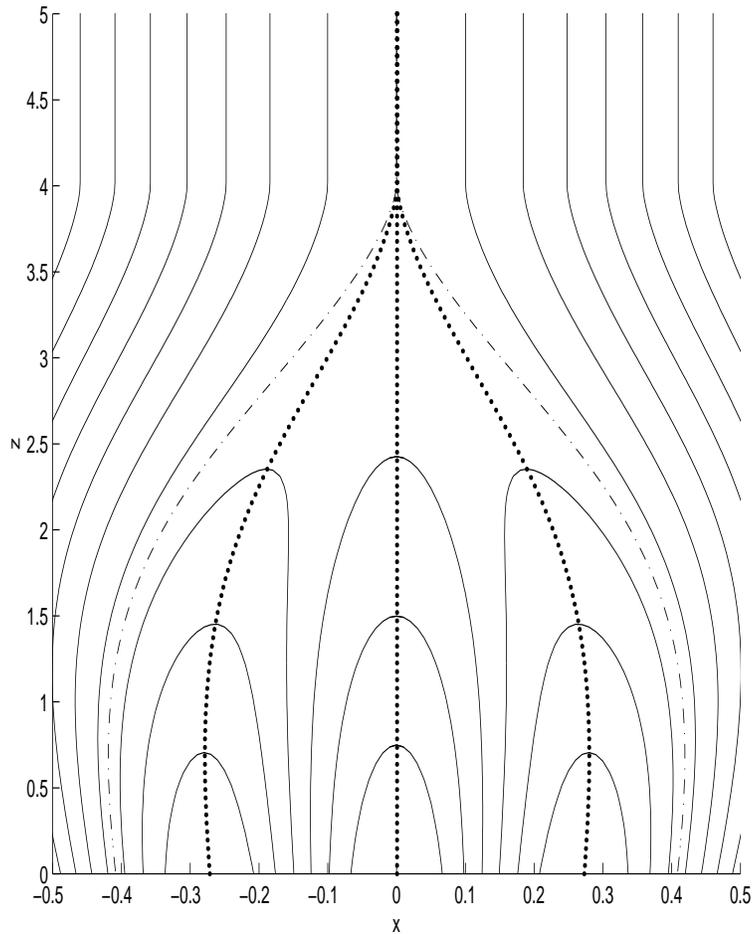}
\caption{Current sheets in helmet streamers.
The dotted lines $\dots$ correspond to the center of each streamer and
the dash-dotted lines to the separatrices between open and closed field
lines. These lines are the locations where thin current sheets
are likely to form.}
\label{fig1}
\end{figure}
In this paper, however, we  show results mainly for
the second resistivity model. The reason is that
one possible mechanism to produce anomalous resistivity
is the interaction of particle and waves. Candidates for such
micro-instabilities are e.g.\ the lower hybrid drift instability and
ion acoustic instability. These instabilities are well known
to occur in regions with large electric current densities, e.g.
thin current sheets.
Figure \ref{fig1} gives an
overview about the location of thin current sheets within our triple
helmet streamer models. Thin current sheets may form in the center
of each streamer, similar to the formation of thin current sheets
in the Earth's magnetotail as investigated in e.g.
\inlinecite{wiegelmann:schindler95}.
Another location for thin current sheets are the separatrices
between the closed field lines of helmet streamers and open
field lines, because the differential rotation will cause magnetic
shear on closed field lines (see {\it paper I} for details).
It is thus reasonable to choose a resistivity profile which localizes
the resistivity at the known locations of the strong currents.
This is the case for the second and the third resistivity model.
Since the difference in the results between both models turned out to be
small in all cases we investigated, we decided to use the second
resistivity model because it is simpler.

Therefore, in most
of the  results presented  (except in Figure \ref{fig7}) we used the
second resistivity model.
We also performed test runs with the constant resistivity model.
They gave results
qualitatively similar  to the other models corroborating
the results of \inlinecite{dissao} and
\inlinecite{otto:etal90}. Details of the resistivity
models can be found in the Appendix.

\section{Results of time-dependent numerical experiments}
\label{results}

\subsection{Configurations without cusp}
\begin{figure}
\includegraphics[height=13cm,width=12cm]{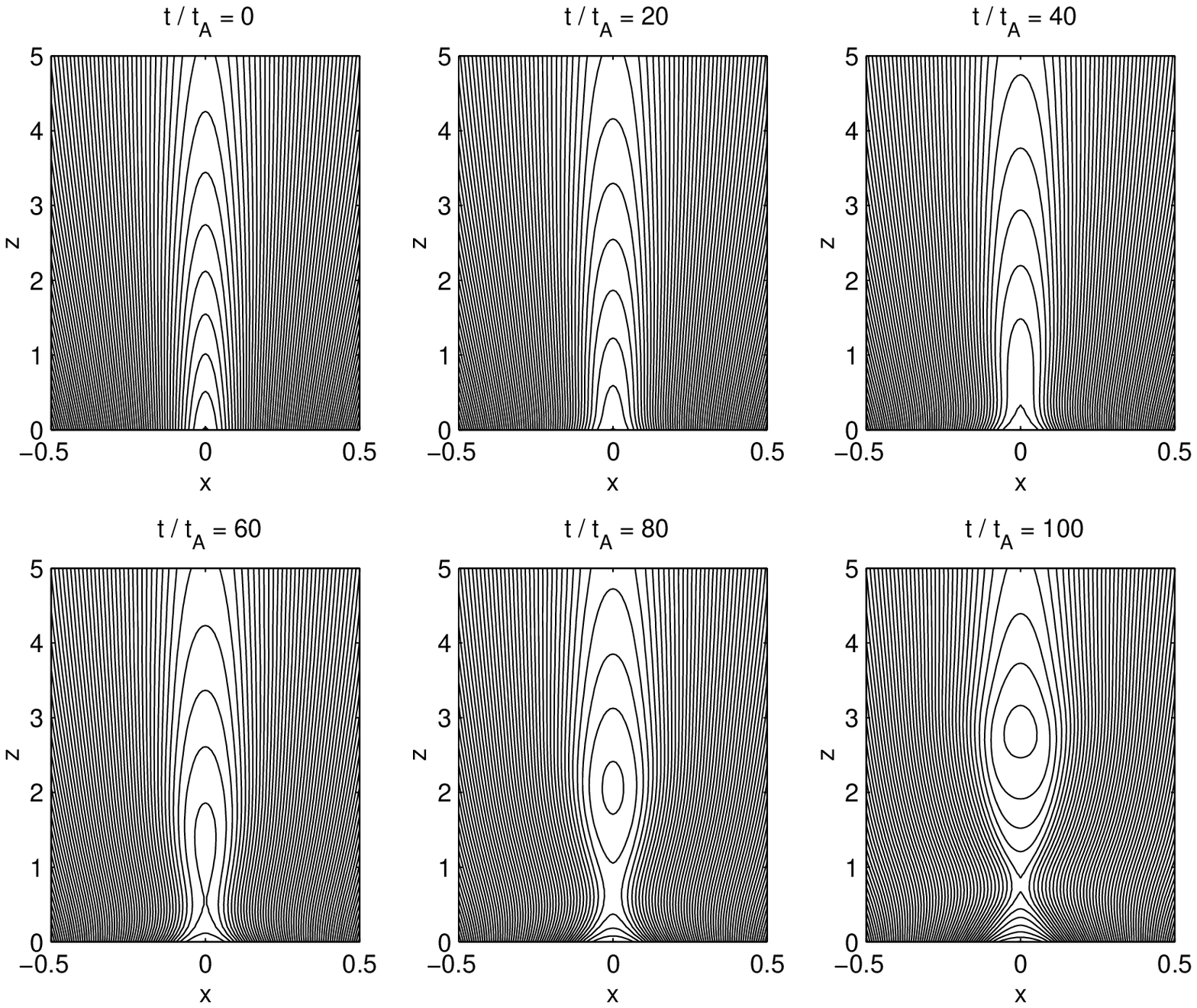}
\caption{time-dependent evolution of magnetic field lines in a single streamer.
The time is measured in units of the Alfv\'en time.}
\label{fig2}
\end{figure}
Before carrying out the numerical experiments
for the  triple streamer configurations,
we performed a numerical experiment for a single
streamer
with similar
 parameter values.
We use this experiment as a reference case. Details of the
equilibrium configuration, the used grid size and the
resistivity model can be found in the Appendix.
Comparison of the reference
case with the triple streamer cases will allow us to work out
particular differences in the dynamical behaviour of the triple
streamers models.
Within our model, a single streamer
is very similar to a model of the Earth's magnetotail and thus our
investigations of a single streamer confirm the well known
results of magnetotail MHD simulations
(e.g. \opencite{birn80}; \opencite{dissao}; \opencite{otto:etal90}).
As shown in Figure
\ref{fig2} the configuration stretches during its evolution and
after some time an X-point forms. A plasmoid appears above this
X-point and is accelerated into interplanetary space.
This
process may be interpreted as a simple model for the development
of a coronal mass ejection. In Figure \ref{fig2} and in all other
figures showing a time evolution the time-scale is given in units
of Alfv\'en
times $t/t_A$.

\begin{figure}
\includegraphics[height=13cm,width=12cm]{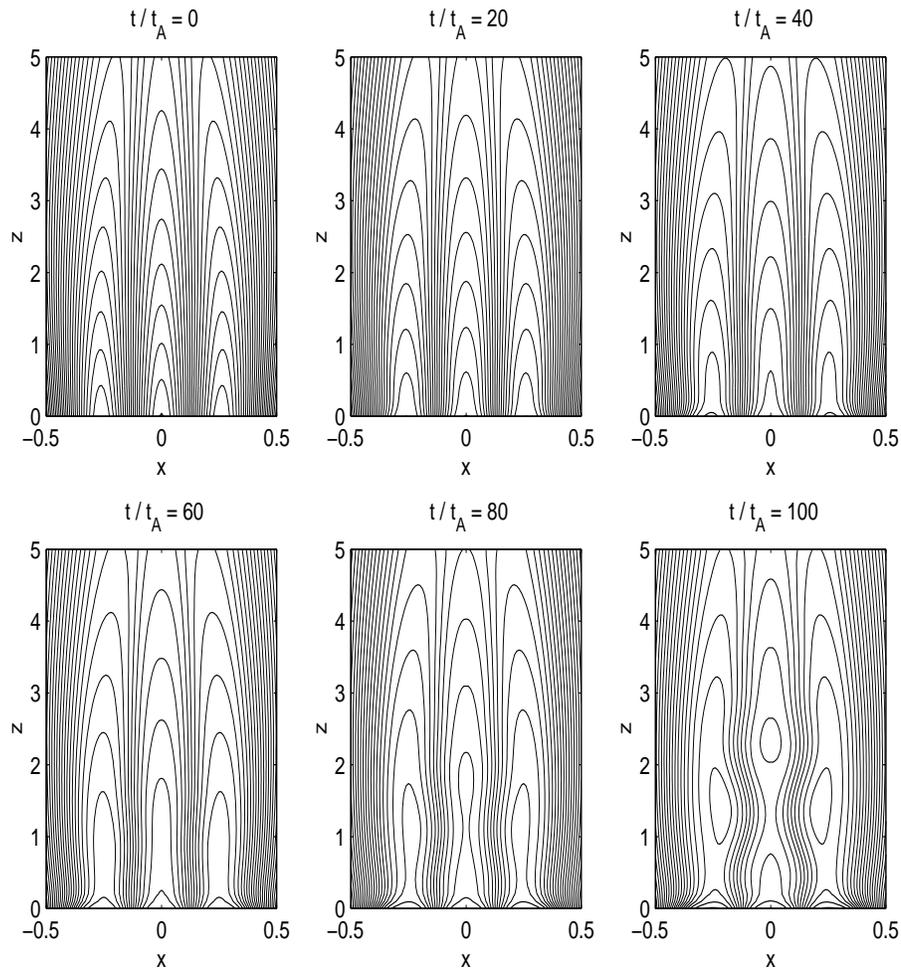}
\caption{Time-dependent evolution of magnetic field lines in triple streamers.}
\label{fig3}
\end{figure}

As the next example we investigate the dynamics
of three parallel helmet streamers. The start configuration
is given by the equilibrium with parameter set {\it a }
given in Table I of {\it paper I} (see also the Appendix).

In Figure \ref{fig3}
we show plots of the field line evolution.
As one can see in Figure
\ref{fig3}, a plasmoid forms in all three streamers. In principle the
process of plasmoid formation and acceleration in each streamer
is similar to the same process in a single streamer.
There are, however, some
differences which we would like to point out. When we compare the
triple streamer evolution with the single streamer evolution,
it is conspicuous that the dynamics of
the middle streamer is slower than the dynamics of the
single streamer in Figure \ref{fig2}.
Furthermore, the X-points forms somewhat higher up
in the corona than in the single streamer.
One also sees that the
plasmoid in the middle streamer forms higher up in the corona
than the plasmoids in the outer streamers. We suggest the following
physical explanation for these differences between
the single and triple streamer cases.

Within the process of magnetic reconnection, plasma and the
frozen-in magnetic field are transported into the
reconnection region. This
leads to a deformation of the surrounding magnetic field outside
the reconnection region as well.
This deformation outside the
reconnection area is not subject to changes in magnetic
topology because the plasma there is frozen-in. The two outer
streamers of the triple structure make this deformation of magnetic
field lines much more difficult than the open field lines outside
a single streamer. In our opinion,
this leads to the observed slower dynamic
evolution of the triple streamer configuration. This stabilizing
effect of the two outer streamers towards the middle streamer is
similar to the well-known boundary stabilization in other
stability problems. One finds that
a boundary consisting of ideally conduction walls, which are impermeable for
plasma and magnetic fields, can have a stabilizing influence
(for an example in the framework of solar physics see
e.g. \opencite{platt:neukirch94}).
If the separation of the boundaries
 parallel to the plasma sheet is small, the
stabilizing effect can be so strong that
no reconnection occurs. In our case the outer streamers of a triple
structure are not as rigid as a conducting wall, but still
much more
rigid than open field lines, thus explaining the slower time evolution.

The fact that the reconnection site in the
middle streamer is located higher up than in a comparable single streamer
is probably again caused by the constraints  imposed
by the two outer
streamers. The plasma flow towards the reconnection site
in any of the outer streamers is only
restricted on one side, while the plasma flow towards the reconnection site
in the
inner streamer is restricted on both sides.
It is therefore no surprise that the X-points in the
outer streamers form approximately at the same height as in a
single streamer. This in turn leads to a deformation of the inner streamer
field lines towards the outer streamers at this height. For the
formation of an X-point at the same height
in the middle streamer it would be
necessary that
the field lines deform towards the center of the middle streamer
and this is just the direction opposite to  their actual
deformation. Thus the
formation of X-points in all three streamers at the same height
is impossible. As a result the
formation of the X-point in the middle streamer
occurs higher up in the corona than in a
similar single streamer. On the other hand, the
middle
streamer in turn imposes geometrical restrictions on the two
outer streamers. These restrictions are, however, comparatively minor and
the X-points in the outer streamers form
somewhat lower down than in a similar single streamer. We attribute this to the
lack of symmetry within  the two outer streamers.

We remark that the described effects are almost independent
of the  resistivity model. For the results shown
in Figure \ref{fig2} and \ref{fig3}
we used the model which localizes the
resistivity in the center of the current sheet inside each streamer.
Simulations with the
constant resistivity model and the current dependent
resistivity model lead to very similar results.

\subsection{Configurations with cusp}

\begin{figure}
\includegraphics[height=13cm,width=12cm]{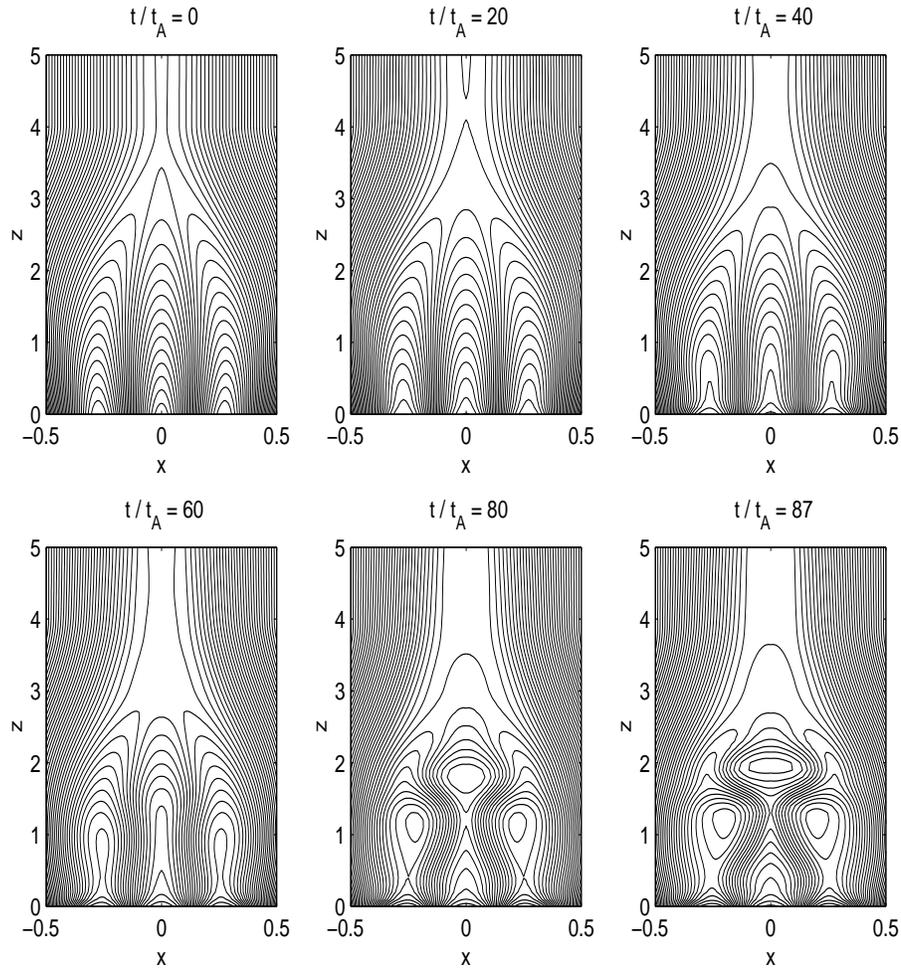}
\caption{Time-dependent evolution of triple streamers with cusp structure.}
\label{fig5}
\end{figure}

Pictures of helmet streamers usually show a typical cusp structure
which is located at the transition from the closed field line
region to the open field line region. In {\it paper I}
we have calculated configurations with a cusp structure which we now
use as initial conditions for our numerical experiment.

Figure \ref{fig5} shows an example of the time-dependent
evolution of a helmet streamer configuration with cusp structure.
The initial condition is given by equilibrium {\it a} in Table II in {\it paper I}
(see also Appendix).
We localized the resistivity at the current sheets in the the center
of each streamer and at the current sheets at the boundary between
open and closed field lines (see Figure \ref{fig1} for an
overview about the current sheet system of triple helmet
streamers).

One finds five regions within the
configuration where magnetic reconnection can occur. We illustrate
these processes schematically in Figure \ref{fig6}:
\begin{figure}
\includegraphics[height=13cm,width=12cm]{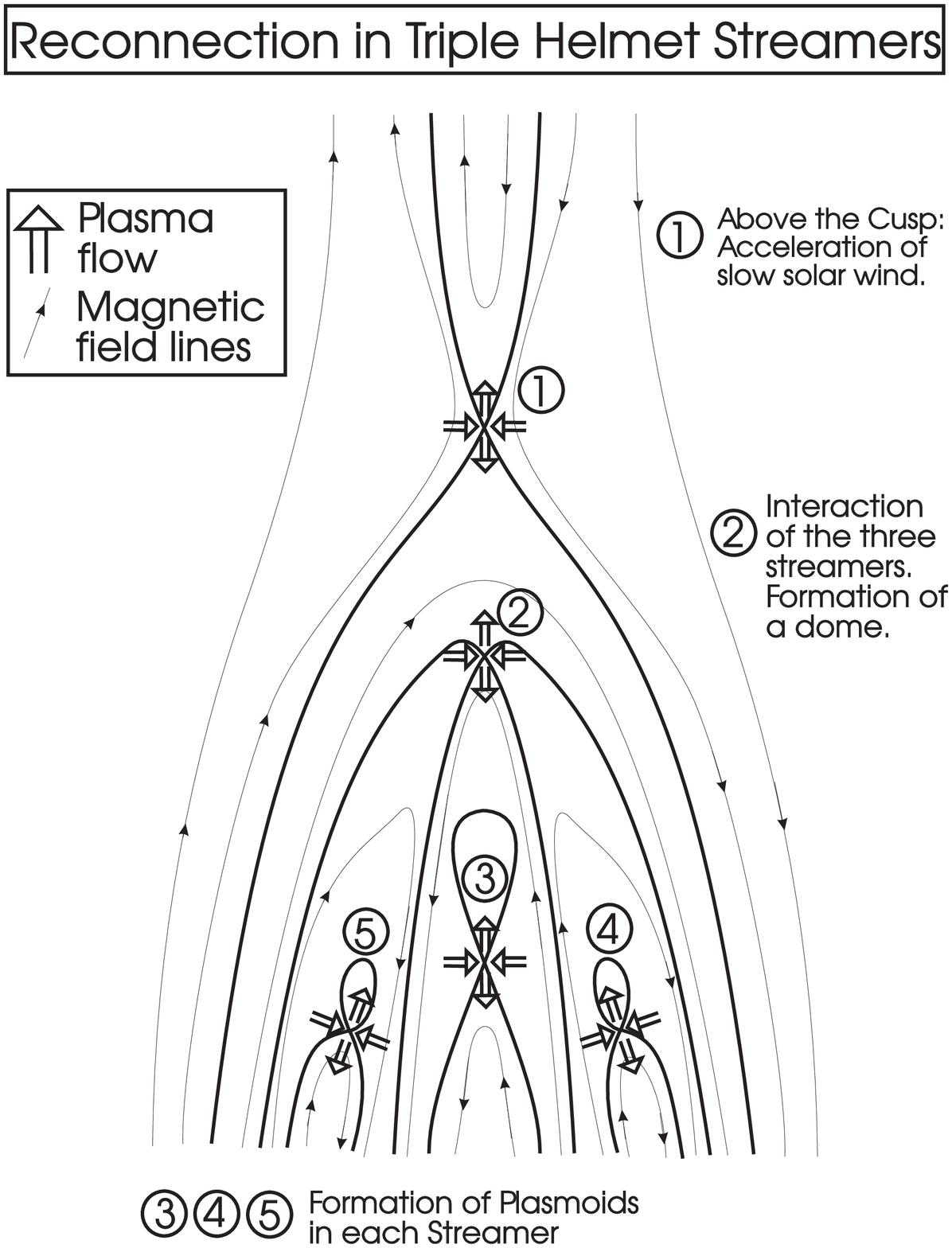}
\caption{Schematic illustration of processes within triple streamers with cusp.}
\label{fig6}
\end{figure}
\begin{itemize}
\item In the reconnection region (1) plasma and the frozen-in
magnetic flux of open field lines are transported into the center
and after a short time an X-point forms. At this X-point
magnetic energy is transformed into kinetic energy and heat. The
reconnected magnetic field above
the cusp is accelerated into
interplanetary space. Below the cusp a dome forms which is
located above the
three streamers. We suggest that
reconnection processes like this one can
be a significant source of plasma and magnetic flux for
the wind emanating from streamer regions.
\item In region (2) somewhat below the cusp we observe
the following
processes.
\begin{itemize}
\item We find interaction between the three streamers and an
X-point forms between the top parts of
the two outer streamers. Above this X-point a
dome forms similarly to the process (1). Below the X-point the
magnetic flux of the middle streamer increases. We call this process
(2a).
\item In principle this process could also happen in the opposite
direction: The dome formed in region (1) is transported
downward by the plasma flow. Simultaneously
plasma of the middle streamer flows upward into the reconnection
region around the X-point and consequently the magnetic flux of the
outer streamers increases. For this process to occur
it is necessary that
the magnetic field configuration in the top part of
the dome is flat, because for
spontaneous reconnection the angle of plasma inflow has to be
larger than the angle in the outflow region. We call this process (2b).
\end{itemize}
\item The processes (3),(4) and (5) may occur similarly in each
streamer. Plasma is transported into the center of each
streamer and an X-point forms in each of these regions.
A plasmoid forms inside the streamers and is accelerated.
Similar to the situation of the three parallel streamers
without cusp structure, the X-point
in the inner streamer forms higher up in the corona than in the outer
streamers.
\end{itemize}
It is  useful to know which
of these processes could occur simultaneously
if the reconnection process was stationary. In
two-dimensional configurations and for stationary magnetic reconnection
it is necessary that the electric current density in the invariant
direction ($j_y$) has the same sign in all reconnection regions (and
thus in whole space). Thus stationary
magnetic reconnection would be possible
simultaneously for the processes (1), (2b), (4), (5) on the one hand
or for the processes (2a), (3) on the other hand.

Independently of the used resistivity profile (constant,
localized, current-dependent resistivity),
process (1) always occurs earlier than
the other processes. As the current density and thus the
resistivity is assumed to be very large in the heliospheric
current sheet above the cusp (see {\it paper I} for details), we also
carried out numerical experiments with
a localized resistivity in this area only. In that case only
process (1) occurs which we tentatively
identify with a possible mechanism of the acceleration of the slow solar wind in the
helmet streamer stalk. The triple streamer configuration
below the cusp does not change very much in this case.
This corresponds to an
approximately static streamer belt, from which plasma and magnetic flux
are continuously expelled. Since we have not included gravitation and
a possible background flow, a final assessment of the relevance of our model
for the slow solar wind is not yet possible.
We remark that
the triple structure of the closed field line region
of our streamer model does
not play a major role for this process so that this process would also occur
above a single streamer. The acceleration process suggested here is
somewhat similar to that investigated of \inlinecite{einaudi:etal99}.
The main differences are that we have not included flow in our equilibrium model,
that we investigate only resistive processes and that our initial conditions
are two-dimensional.

\begin{figure}
\includegraphics[height=13cm,width=12cm]{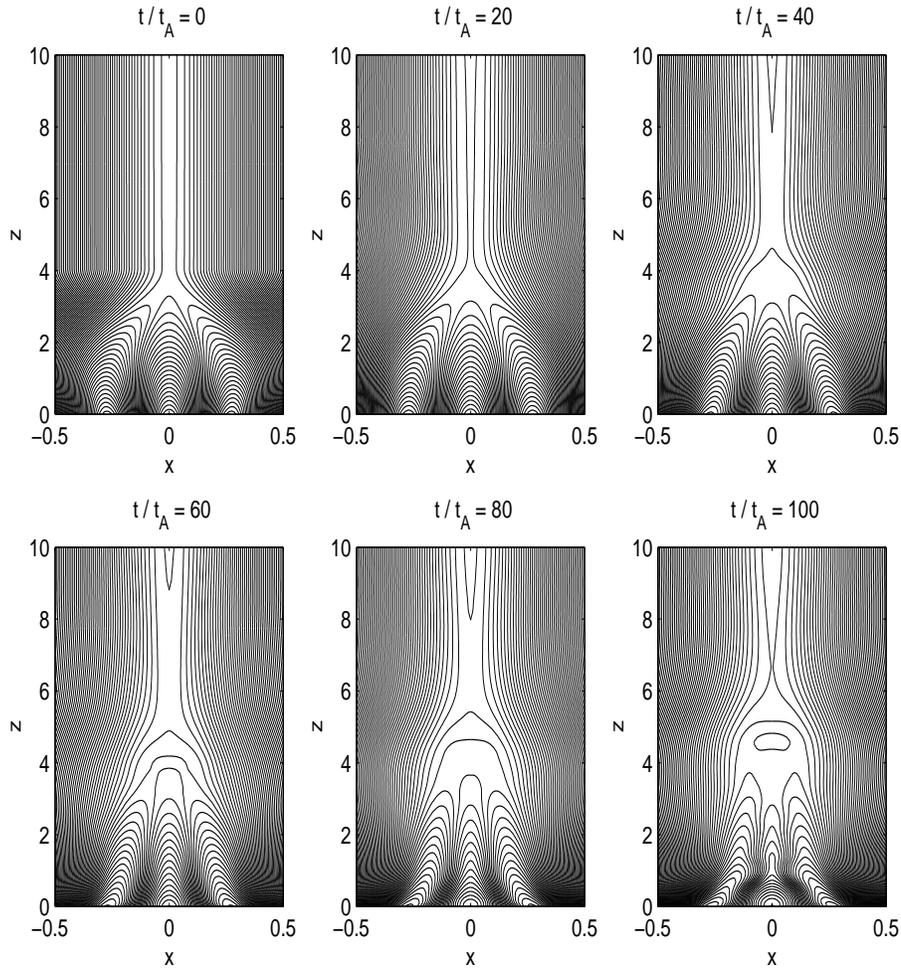}
\caption{Time evolution of triple helmet streamers with a different
resistivity model. (note the different scales in $z$ compared with the other pictures)}
\label{fig7}
\end{figure}
For the configurations with cusp we also find that the evolution
of the middle streamer is slower as compared to the
evolution of a single streamer, very similar to our results for
three parallel streamers without cusp. We find, however, that
for the configuration with cusp
the dome formed by the processes
(1) and (2) has the additional effect that the plasmoid formed
in the middle streamer cannot leave the configuration
(and pass through the dome) without problems. The reason is
that the dome has the wrong magnetic polarity to allow the plasmoid
to pass by further reconnection.
The plasmoid which is accelerated in
process (3) is pushed against the cusp and additional
reconnection processes are necessary to let it pass through the dome
by processes in analogy
to the process (2b).
So far, however, we have not been able to find this process
within our numerical experiments.
One possible explanation for this
is that the dome has the possibility to move
outward completely without any reconnection process. The results of the numerical
experiments presented in Figure \ref{fig7} with a current dependent
resistivity and a decreasing density profile $(\rho=p \exp (-\frac{z}{H})$
show some indications
for this rise of the whole configuration including the dome.
The X-point of region (1) is first located  at $z=4$ and then rises
to $z=7$ in the last snapshot of Figure \ref{fig7}. This slow
rise of the dome seems to be  similar to the slow coronal
mass ejections observed with LASCO on SOHO \cite{schwenn99, nandita}
where the
magnetic field lines are connected with the sun for a considerable
longer time than other CME's.

A remark is necessary
concerning open and closed field
lines in multiple streamer structures.
In the case of parallel streamers we defined
magnetic field lines as open if they cross the upper boundary.
In the case of cusp solutions
open field lines are outside the cusp separatrix
and closed field lines inside
(see {\it paper I} for details).
In the case of three parallel
streamers, open field
lines exist between the streamers.
This is not the case for for triple streamers with cusp
structure. Thus configurations with cusp structure have a different
magnetic topology than those without a cusp. It seems
interesting to ask which of
these cases is closer to reality. On the one hand, as we mentioned above,
the observations often give the impression of a cusp-like structure for
helmet streamers but on the other hand new observations
\cite{inhester98} show that
within the extended streamer belt localized regions with open
field lines exist. We conclude that within our two-dimensional theory
we cannot model cusp-like structures and
open field line regions at the same time. This shortcoming can
only be overcome by a future three-dimensional model.

\section{Conclusions and Outlook}
\label{conclusions}
In this paper we have
tried to make a step
towards a better
theoretical understanding of the dynamics of helmet streamers
with triple structure. We investigated the possible role of
triple streamers for the development of
coronal mass ejections and
as a possible source for plasma and magnetic field for the wind
emanating from streamer regions.
In previous works
(e.g. \opencite{steinolfson94}; \opencite{linker94};
\opencite{wu:etal95}; \opencite{wu97}
)
only single helmet streamer where
modelled and these models assumed the slow solar wind as a stationary
plasma flow on open field lines in the streamer region. To get a
start equilibrium these authors solved the ideal time-dependent MHD
equations numerically until a stationary state was reached. These
works
showed that helmet streamers can become unstable and produce
coronal mass ejections.

The  present
investigations where motivated by the new observations with the
LASCO coronagraph on SOHO \cite{schwenn97}. These
observations showed, that the
streamer belt in the solar activity minimum typically has a triple structure.
The observations also gave further
strong evidence that a stationary
slow solar wind may not exist but is produced by many small
eruptions.
Apart from these continuously occuring small eruptions, also
large, however rarely occuring
coronal mass ejections are generated in the triple streamer belt.

To take
these observations into account in a helmet streamer model, we
developed an analytical stationary  model of triple
helmet streamers using the ideal MHD equations in {\it paper I}.
The initial states have to possess a non-vanishing free energy to allow
their instability with respect to magnetic reconnection. As discussed in {\it paper I}
we took into account the observation that the streamer configurations are very extended in the radial
direction to simplify the calculation of the initial states. We emphasize that  such radially
extended configurations cannot be modeled by potential fields.
In the present paper we investigated the stability of these
stationary state configurations with the help numerical experiments
in the framework of time-dependent resistive MHD.
We used three ad hoc models for the resistivity,
a constant resistivity, a resistivity localized
at the thin current sheets
and a current-dependent
resistivity.
We first investigated the ideal stability of our triple
streamer configurations and found that they are stable on
the time-scale of our simulations.

Next we investigated the resistive stability of a triple
streamer
configuration without cusp structure.
We found that our triple streamer configuration
is resistively unstable. Reconnection takes place
and
plasmoids form
in each of the three closed field line regions.
By comparing the time evolution
of the triple streamer model with a single streamer model
we found that for a triple streamer configuration without
cusp structure the time-evolution is usually slower than
the corresponding time evolution of a single streamer. The triple
streamer evolution also shows characteristic differences
in comparison with the single streamer case concerning the location of the
reconnection sites. We could explain these differences by the influence the
three streamers exert on each other.

For triple streamer configurations with cusp structure we found
quite similar results for reconnection processes inside the
closed field line regions but in addition  we
found that the helmet streamer stalk above the cusp is highly
unstable to reconnection. This reconnection
process leads to the formation of a dome above the triple structure,
i.e. a region of closed field lines which encloses the triple structure
completely. The resistive instability of the
streamer stalk current sheet could be a possible source of plasma
and magnetic field for
the non-steady solar wind emanating from the streamer regions.
In the present paper we have only been able to demonstrate that
this mechanism works with a preexisting cusp structure. In later
stages of our simulations the cusp is replaced by an X-point at
which reconnection can take place. The inclusion of flow on open
field lines would also allow for the possibility to generate a
new cusp structure making a repetition of the process possible.

Furthermore we found   interaction between the
two outer streamers just below the cusp region. This interaction
can also contribute to the formation
of the dome. This dome
makes it more difficult for plasmoids to escape and thus streamers
with cusp structure within our model
are less likely to eject material
than the configurations without cusp.

These results are consistent  with the observational
finding that the triple streamer configuration is observed
to be stable for several days.
One may also speculate
about the fact that the observations
usually show three streamers which approximately have
the same radial extension.
A possible explanation on the basis
of our model is that if one of the streamers
grows and becomes much larger than the other streamers,
it becomes prone to instability and a coronal
mass ejection occurs
similarly to the case of a single streamer.
In this process the streamer looses energy, mass and magnetic flux
and returns to its original state.

The numerical experiments presented here can only be considered
as a very first step towards a complete model of these interesting
phenomena. We already mentioned above that although the models with
cusp structure seem to be matching the observed streamer structure
best of all our models, it is not possible to include regions
of open field lines  between the streamers in our models. One way
to overcome this shortcoming would be to use three-dimensional models,
which is a natural next step.
Other possible improvements of the present work are the inclusion of
gravity and the use of spherical geometry.

\acknowledgements

It is a pleasure to
thank Lutz Rast\"atter for supplying
the numerical code used in this paper and for
his help with numerical problems. We also thank J\"urgen Dreher
and Andreas Kopp for useful discussions of the numerical aspects
of this paper
and Bernd Inhester,
Rainer Schwenn and Nandita Srivastava
 for sharing their knowledge of the LASCO observations
with us. One of us (TN) is grateful to PPARC for financial support by an Advanced
Fellowship.

\appendix

In this section we briefly list the  model parameters used for
the initial streamer configurations shown in the figures and
details about the  grid sizes and the resistivity models.

In Figure \ref{fig2} we used the model parameters $s_1=0.8,
s_2=0.4, s_3=0.2$, $c_1=15$ for the initial conditions (we refer the
reader to
{\it paper I}, Section 3.1 for a definition of these parameters).
This configuration corresponds to the middle streamer in
{\it paper I}, Figures 2a and 2b. In Figure \ref{fig3}
we used as initial conditions
the configuration  a given in {\it paper I}, Table I and
shown in Figure 2a of {\it paper I}.
In Figure \ref{fig5} we used as initial conditions
the configuration a given
in {\it paper I}, Table II and shown in Figure 3a of {\it paper I}.

We used a grid of $53$ points in $x$ and $z$ in Figures \ref{fig2},
\ref{fig3}, \ref{fig5} and
a grid of $53$ points in $x$ and $105$ points in $z$ in Figure \ref{fig7}.
The grid is rectangular and we only calculated one half of each
configuration ($x=0 \dots 0.5$) and get the other half of the
configurations ($x=-0.5 \dots 0$) by symmetry.
The coordinate $z$ runs from $0 \dots 5$ in Figures \ref{fig2},
\ref{fig3}, \ref{fig5} and from
$0 \dots 10$ in Figure \ref{fig7}.

In Figures \ref{fig2}, \ref{fig3} and \ref{fig5}
we used a resistivity profile  localized at the
equilibrium current sheets shown in Figure \ref{fig1}. The current
sheets are located in the center of each streamer and at the boundary
between open and closed field lines.
Thus the position of the current sheets $x_{ss}(z)$ is calculated
analytically as described in {\it paper I} and the resistivity
profile was chosen as
$\eta = \eta_0 \exp \left(-(x-x_{ss})^2/{b}\right)$ with
$\eta_0 = 0.001, b=20$. In Figure \ref{fig7} we used a current dependent
resistivity profile in the form $\eta=\eta_0 \exp(-a (z-z_{\mbox{cusp}}))
\mbox{tanh}(|\vec j|)$ with $\eta_0=0.0005$,
$a=0.1$ and $z_{\mbox{cusp}}=4$.

We mention that due to the finite grid size, there will always be small
numerical fluctuations present which are superposed onto the smooth initial
conditions given by the ideal equilibria. The full initial conditions are
therefore given by a smooth component plus a small fluctuating part. We
emphasize that we did not start the instability by adding an explicit
finite amplitude perturbation to the ideal equilibria.

%

\end{article}


\begin{thebibliography}{}
\bibitem[\protect\citeauthoryear{Antiochos et al.}{1999}]{antiochos:etal99}
Antiochos, S.K., DeVore, C.R. and Klimchuk, J.A.: 1999,
{\it Astrophys. J.} {\bf 510}, 485.

\bibitem[\protect\citeauthoryear{Bavassano et al.}{1997}]{bavassano:etal97}
Bavassano,  B., Woo, R.\ and Bruno, R.: 1997
{\it Geophys. Res. Lett.} {\bf 24}, 1655.

\bibitem[\protect\citeauthoryear{Birk et al.}{1997}]{birk:etal97}
Birk, G.T., Konz, C., and Otto, A.: 1997, {\it Phys. Plasmas\/}
{\bf 4}, 4173.

\bibitem[\protect\citeauthoryear{Birn}{1980}]{birn80}
Birn, J.: 1980,
{\it J. Geophys. Res.}
{\bf 85}, 1214.

\bibitem[\protect\citeauthoryear{Biskamp and Welter}{1989}]{biskamp:welter89}
Biskamp, D. and Welter, H.: 1989,
{\it Solar Phys.} {\bf 120}, 49.

\bibitem[\protect\citeauthoryear{Crooker et al.}{1993}]{crooker:etal93}
Crooker, N.U., Siscoe, G.L., Shodan, S., Webb, D.F.,
Gosling, J.T. and Smith, E.J.: 1993,
{\it J. Geophys. Res.} {\bf 98}, 9371.


\bibitem[\protect\citeauthoryear{Cuperman et al.}{1990}]{cuperman:etal90}
Cuperman, S., Ofman, L. and Dryer, M.: 1990,
{\it Astrophys. J.} {\bf 350}, 846.


\bibitem[\protect\citeauthoryear{Cuperman et al.}{1992}]{cuperman:etal92}
Cuperman, S., Detman, T.R., Bruma, C. and Dryer, M.: 1992,
{\it Astron. Astrophys.} {\bf 265}, 785.

\bibitem[\protect\citeauthoryear{Cuperman et al.}{1995}]{cuperman:etal95}
Cuperman, S., Bruma, C., Dryer, M.  and Semel, M.: 1995,
{\it Astron. Astrophys.} {\bf 299}, 389.


\bibitem[\protect\citeauthoryear{Dahlburg and Karpen}{1995}]
{dahlburg:karpen95}
Dahlburg, R.B. and Karpen, J.T.: 1995,
{\it J. Geophys. Res.} {\bf 100}, 23489.

\bibitem[\protect\citeauthoryear{Dreher}{1997}]{dissjd}
Dreher, J.: 1997,
PhD Thesis, Ruhr-Universit\"at Bochum.

\bibitem[\protect\citeauthoryear{Einaudi et al.}{1999}]{einaudi:etal99}
Einaudi, G., Boncinelli, P., Dahlburg, R.B. and Karpen, J.T.: 1999,
{\it J. Geophys. Res.} {\bf 104}, 521.

\bibitem[\protect\citeauthoryear{Habbal et al.}{1997}]{habbal98}
Habbal, S.R., Woo, R., Fineschi, S., Neal, R.O., Kohl, J., Noci, G.
and Korendyke, C.: 1997, {\it Astrophys. J. Lett.} {\bf 489}, L103.



\bibitem[\protect\citeauthoryear{Hundhausen}{1999}]{hundhausen99}
Hundhausen, A.J.: 1999,
in K. Strong, J. Saba, B. Haisch and J. Schmelz (eds.),
{\it The Many Faces of the Sun}, Springer, 143.

\bibitem[\protect\citeauthoryear{Inhester}{1998}]{inhester98}
Inhester, B.: 1998,
personal communication.

\bibitem[\protect\citeauthoryear{Koutchmy and Livshits}{1992}]
{koutchmy:livshits92}
Koutchmy, S. and Livshits, M.: 1992,
{\it Space Sci. Rev.} {\bf 61}, 393.


\bibitem[\protect\citeauthoryear{Linker and Mikic.}{1995}]{linker94}
Linker, J.A. und Mikic, Z.: 1995,
{\it ApJ}{\bf 438}, L45-L48

\bibitem[\protect\citeauthoryear{Mikic et al.}{1988}]{mikic:etal88}
Mikic, Z., Barnes, D.C. and Schnack, D.D.: 1988,
{\it Astrophys. J.} {\bf 328}, 830.



\bibitem[\protect\citeauthoryear{Noci et al.}{1997}]{noci}
Noci, G., Kohl, J.L., Antonucci, E. Tondello, G., Huber, M.C.E. et al.: 1997
{\it Proceedings of Fifth SOHO Workshop }
 ESA SP-404, 75.


\bibitem[\protect\citeauthoryear{Otto}{1987}]{dissao}
Otto, A.: 1987,
PhD Thesis, Ruhr-Universit\"at Bochum.

\bibitem[\protect\citeauthoryear{Otto and Birk}{1992}]{otto:birk92}
Otto, A. and Birk, G.T.: 1992,
{\it Phys. Fluids B} {\bf 4}, 3811.

\bibitem[\protect\citeauthoryear{Otto et al.}{1990}]{otto:etal90}
Otto, A., Schindler, K. and Birn, J.: 1990,
{\it J. Geophys. Res.} {\bf 95}, 15023.

\bibitem[\protect\citeauthoryear{Platt and Neukirch}{1994}]{platt:neukirch94}
Platt, U. and Neukirch, T.: 1994,
{\it Solar Phys.} {\bf 153}, 287.

\bibitem[\protect\citeauthoryear{Pneuman and Kopp}{1971}]{pneuman:kopp71}
Pneuman, G.W. and Kopp, R.A.: 1971,
{\it Solar Phys.} {\bf 18}, 258.

\bibitem[\protect\citeauthoryear{Rast\"atter}{1997}]{disslr}
Rast\"atter, L.: 1997,
PhD Thesis, Ruhr-Universit\"at Bochum.

\bibitem[\protect\citeauthoryear{Rast\"atter and Neukirch}{1997}]{lrtn}
Rast\"atter, L. and Neukirch, T.: 1997,
{\it Astronomy and Astrophysics} {\bf 323}, 923.

\bibitem[\protect\citeauthoryear{Schindler et al.}{1983}]{ks83}
Schindler, K., Birn, J. and Janicke, L.: 1983,
{\it Solar Phys.}
{\bf 87}, 103.

\bibitem[\protect\citeauthoryear{Schwenn et al.}{1997}]{schwenn97}
Schwenn, R., Inhester, B., Plunkett, S.P., Epple, A., Podlipnik, B.,
et al.: 1997,
{\it Solar Phys.}, {\bf 175}, 667.

\bibitem[\protect\citeauthoryear{Schwenn}{1999}]{schwenn99}
Schwenn, R.: 1999,
{\it Proceedings Plastro 98}, in press.

\bibitem[\protect\citeauthoryear{Srivastava et al.}{1999}]{nandita}
Srivastava, N., Schwenn, R., Inhester, B., Stenborg G., and Podlipnik, B.,
{\it Solar Wind Nine, AIP CP 471}
 edited by S. R. Habbal, R. Esser, J. V. Hollweg
and P. A. Isenberg, pp. 115-118, New York, 1999

\bibitem[\protect\citeauthoryear{Steinolfson}{1994}]{steinolfson94}
Steinolfson, R.S.: 1994,
{\it Space Sci. Rev. }{\bf 70}, 289.

\bibitem[\protect\citeauthoryear{Wang et al.}{1993}]{wang:etal93}
Wang, A.-H., Wu, S.T., Suess, S.T. and Poletto, G.: 1993,
{\it Solar Phys.} {\bf 147}, 55.

\bibitem[\protect\citeauthoryear{Wang et al.}{1997}]{wang:etal97}
Wang, S., Liu, Y.F. and Zheng, H.N.: 1997,
{\it Solar Phys.} {\bf 173}, 409.

\bibitem[\protect\citeauthoryear{Wiegelmann and Schindler}{1995}]
{wiegelmann:schindler95}
Wiegelmann, T. and Schindler, K.: 1995,
{\it Geophys. Res. Lett.} {\bf 22}, 2057.

\bibitem[\protect\citeauthoryear{Wiegelmann et al.}{1998}]{paper1}
Wiegelmann, T., Schindler, K. and Neukirch, T.: 1998, {\it Solar
Phys.}{\bf 180}, 439.

\bibitem[\protect\citeauthoryear{Woo et al.}{1995}]{woo:etal95}
Woo, R., Armstrong, J.W., Bird, M.K. and P\"atzold, M.: 1995,
{\it Astrophys. J.} {\bf 449}, L91.

\bibitem[\protect\citeauthoryear{Wu et al.}{1995}]{wu:etal95}
Wu, S.T., Guo, W.P. and Wang, J.F.: 1995, {\it Solar Phys.} {\bf
157}, 325.

\bibitem[\protect\citeauthoryear{Wu and Guo}{1997}]{wu97}
Wu, S.T., Guo, W.P.: 1997,
in {\it Coronal Mass Ejections: Causes and Consequences}
(N. Crooker, J. Joselyn and J. Feynman eds.)
AGU Geophysical Monograph Series, American Geophysical Union,
Washington, DC, 1997

\bibitem[\protect\citeauthoryear{Yan et al.}{1994}]{yan:etal94}
Yan, M., Otto, A., Muzzell, D. and Lee, L.C.: 1994,
{\it J. Geophys. Res.} {\bf 97}, 16789.

\end{thebibliography}
\end{document}